\documentclass[12pt]{iopart}

\usepackage{graphicx}
\begin{document}

\title[STAR results from beam energy scan program]{STAR experiment results from the beam energy scan program at RHIC}

\author{Bedangadas Mohanty$^{a}$ for the STAR Collaboration}

\address{$^{a}$ Variable Energy Cyclotron Centre, 1/AF Bidhan Nagar, Kolkata - 700064, India}
\ead{$^{a}$bmohanty@vecc.gov.in}
\begin{abstract}

We present the first results using the STAR detector from the Beam Energy Scan (BES) program 
at the Relativistic Heavy-Ion Collider (RHIC). In this program, Au ion collisions at center 
of mass energies ($\sqrt{s_{\mathrm {NN}}}$) of 7.7, 11.5 and 39 GeV allowed RHIC to extend the study
of the QCD phase diagram from baryonic chemical potential values of 20 MeV to about 400 MeV. 
For the high net-baryon density matter at midrapidity, formed in these collisions, 
we report several interesting measurements. These include the observation of 
difference between anti-particle and particle elliptic flow, 
disappearance of the difference in dynamical azimuthal correlations 
with respect to event plane between same and opposite signed charged particles, 
change in slope of eccentricity at freeze-out and directed flow 
of protons as a function of $\sqrt{s_{\mathrm {NN}}}$ and the deviation of higher order 
fluctuations from hadron resonance gas and Poissonian expectations. Possible interpretations
of these observations are also discussed.
\end{abstract}


\section{Introduction}
One of the main goals of the heavy-ion collision program at RHIC is to study various
aspects of the QCD phase diagram~\cite{pd}. The results from Au+Au collisions at 
$\sqrt{s_{\mathrm {NN}}}$ = 200 GeV have established the existence of a system
with partonic degrees of freedom~\cite{starwp}. Lattice QCD calculations at zero 
baryon chemical potential ($\mu_{\mathrm B}$) show that the quark-hadron transition
is a cross over~\cite{latticenature}. Some of the remaining goals at RHIC are
to search for the signals of phase boundary and critical point (CP)~\cite{bm09}. 
These form the main motivations for the RHIC BES program. In pursuit of which, it 
was first established that RHIC (both accelerator and experiments) can 
operate below injection energies of 19.6 GeV through a test run of 
Au+Au collisions at $\sqrt{s_{\mathrm {NN}}}$ = 9.2 GeV~\cite{starbesprc}. 
Further, a set of observable for the physics goals were formulated~\cite{observablebes}. 
This allowed for the starting of the first phase of the BES program in the year 2010. 
The STAR experiment, with full azimuthal coverage, large and uniform acceptance for
all midrapidity hadrons across all $\sqrt{s_{\mathrm {NN}}}$ is ideally suited for this program.
So far we have collected about 5, 11 and 170 million good events for 
7.7, 11.5 and 39 GeV Au+Au collisions respectively, in addition to taking 
data at higher energies of 62.4 and 200 GeV. Here we present results 
on (a) turn-off of observations
related to partonic degrees of freedom, through the measurements of $v_{\rm 2}$ and
dynamical charge correlations, (b) search for the signatures of softening
of Equation Of State (EOS), through HBT and $v_{\rm 1}$ measurements 
and (c) search for signatures of CP through fluctuation measurements. 

The data presented are at midrapidity from the STAR Time Projection chamber (TPC) 
and Time Of Flight (TOF) detectors~\cite{stardetectors}. A high level trigger was used 
in the BES program to reject background events arising from beam and beam-pipe 
interactions~\cite{hltposter}. The centrality selection was carried out using the 
uncorrected charge particle multiplicity measured in TPC for pseudorapidity,
$\mid \eta \mid < 0.5$~\cite{starbesprc}. 
The particle identification was done by measuring the specific ionization 
energy loss in TPC and the particle velocities using TOF as a function of momentum. 

\begin{figure}
\begin{center}
\includegraphics[scale=0.25]{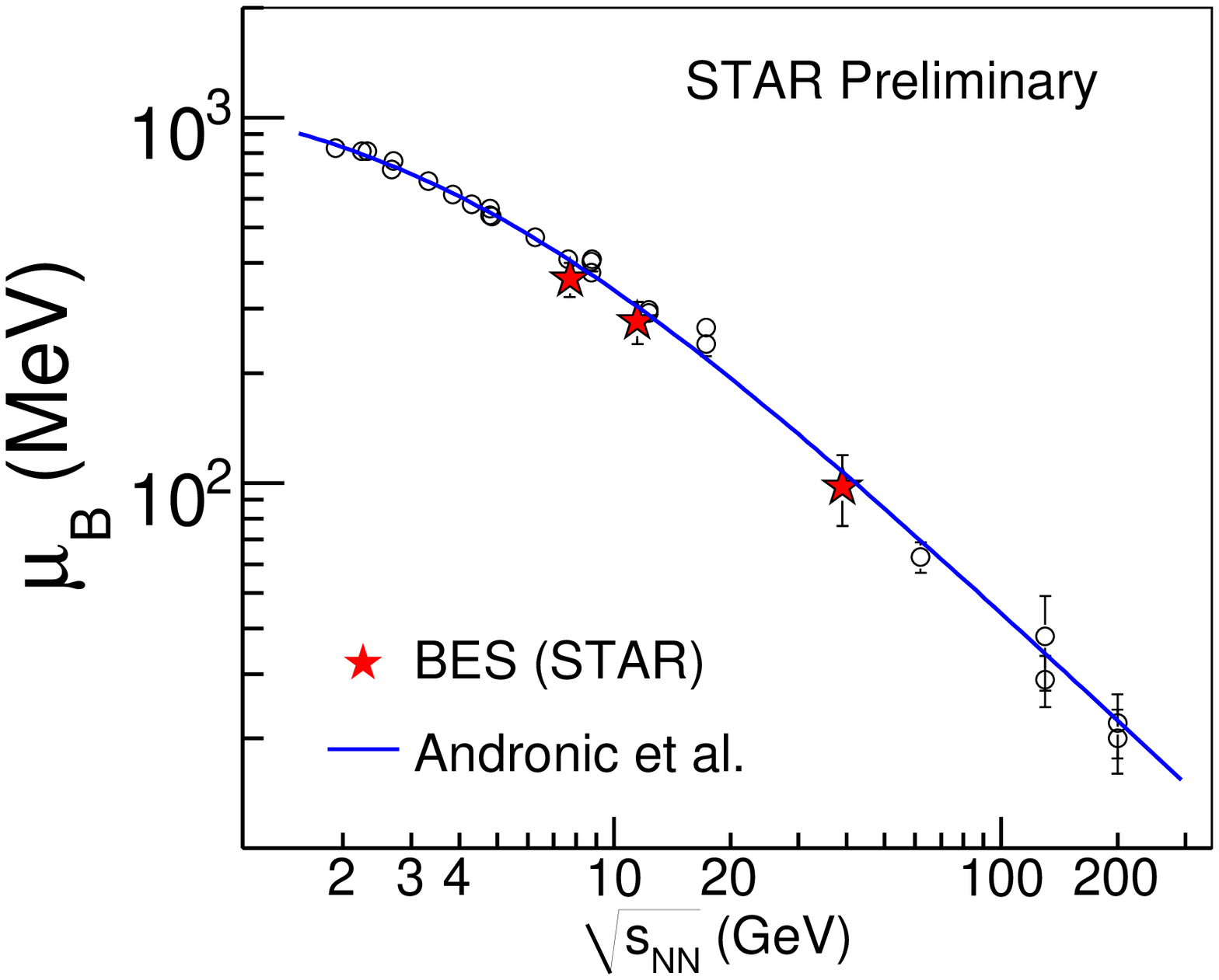}
\includegraphics[scale=0.25]{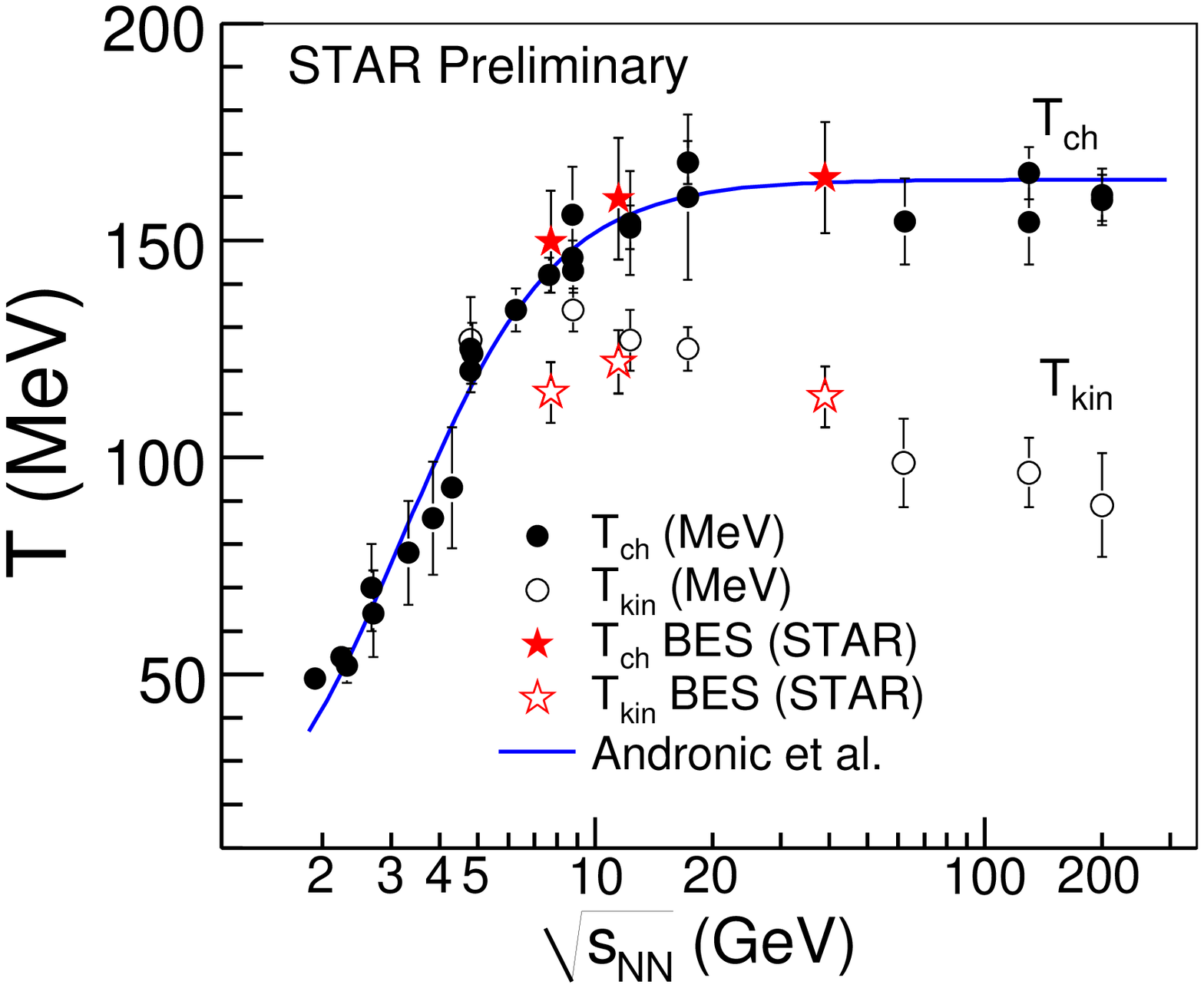}
\includegraphics[scale=0.25]{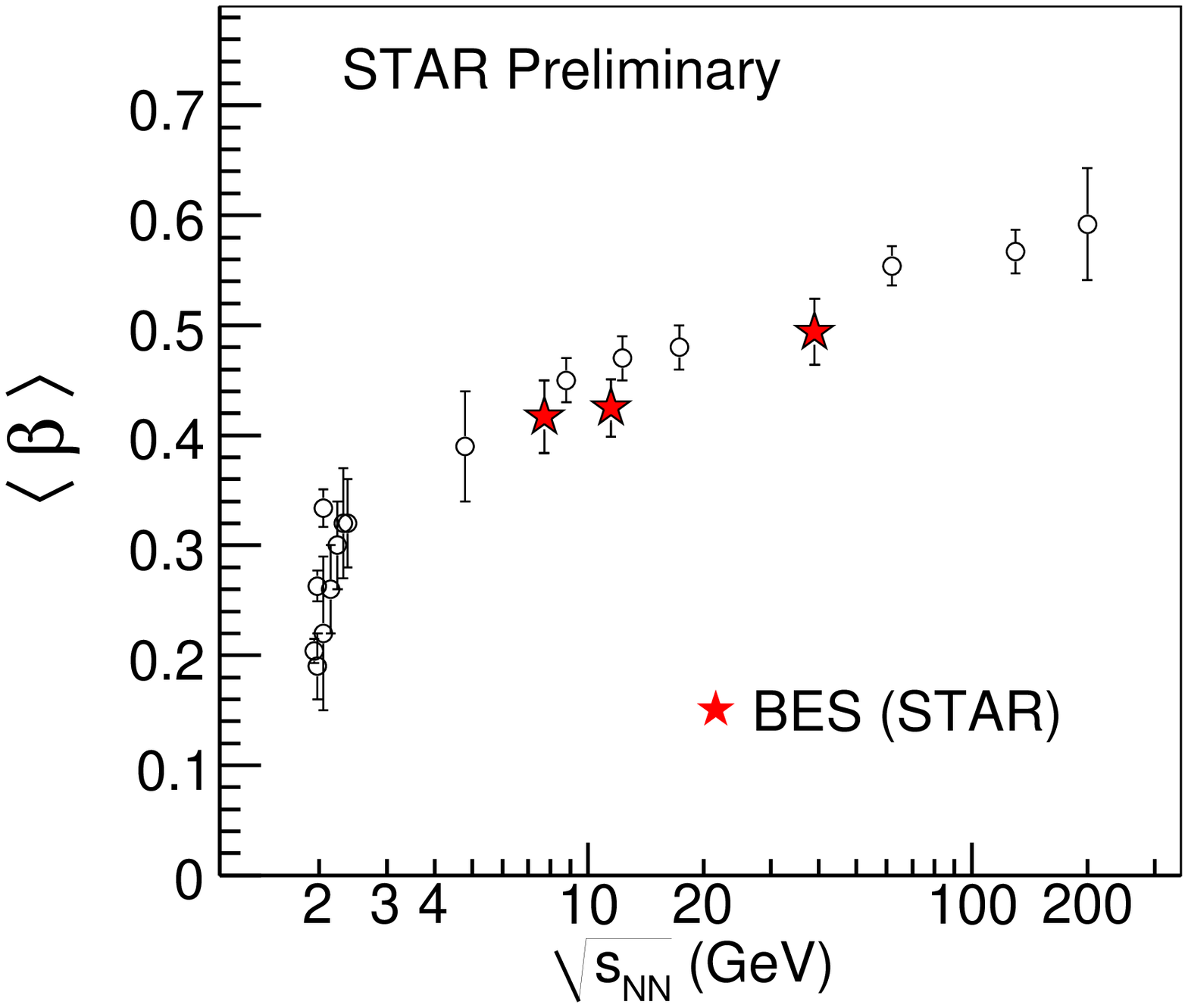}
\vspace{-0.5 cm}
\caption{(Color online) Left panel: $\mu_{\mathrm B}$ vs. $\sqrt{s_{\mathrm {NN}}}$.
Middle panel: $T_{\mathrm {ch}}$ and $T_{\mathrm {kin}}$ vs. $\sqrt{s_{\mathrm {NN}}}$. 
Right panel: $\langle \beta \rangle$ vs.$\sqrt{s_{\mathrm {NN}}}$. The red stars
are new STAR results at midrapidity from 0-5\% central Au+Au collisions 
at BES energies~\cite{lokesh} and the solid line are calculations from Ref.~\cite{andronic}.}
\label{fo}
\end{center}
\end{figure}

Within the framework of a statistical model assuming theromodynamical equilibrium~\cite{lokesh} 
the measured particle ratios ($\pi^{-}/\pi^{+}$,
$K^{-}/K^{+}$, $\bar{p}/p$, $K^{-}/\pi^{-}$ and $\bar{p}/\pi^{-}$) in central
(0-5\%) Au+Au collisions are used to extract the chemical freeze-out (vanishing 
inelastic collisions) conditions. Figure~\ref{fo} shows that the BES program has extended 
the $\mu_{\mathrm B}$ range at RHIC from around 20 MeV to about 400 MeV. The chemical 
freeze-out temperature ($T_{\mathrm {ch}}$) slightly increases with energy from 150 MeV to 
around 165 MeV. The invariant yields of $\pi^{-}$, $K^{-}$ and $\bar{p}$ 
were fitted using a blast wave model to extract the kinetic
freeze-out (vanishing elastic collisions) conditions~\cite{lokesh}. 
The kinetic freeze-out temperature ($T_{\mathrm {kin}}$) is observed to slightly decrease 
whereas the collective radial flow velocity 
($\langle \beta \rangle$) increases as the beam energy increases. The large $\mu_{\mathrm {B}}$ values 
at midrapidity indicates formation of a high net-baryon density matter and which is expected to
reach a maximum value around 7.7 GeV~\cite{randrup}.

\section{Partonic versus hadronic degrees of freedom}

\subsection{Elliptic flow}
\begin{figure}
\begin{center}
\includegraphics[scale=0.3]{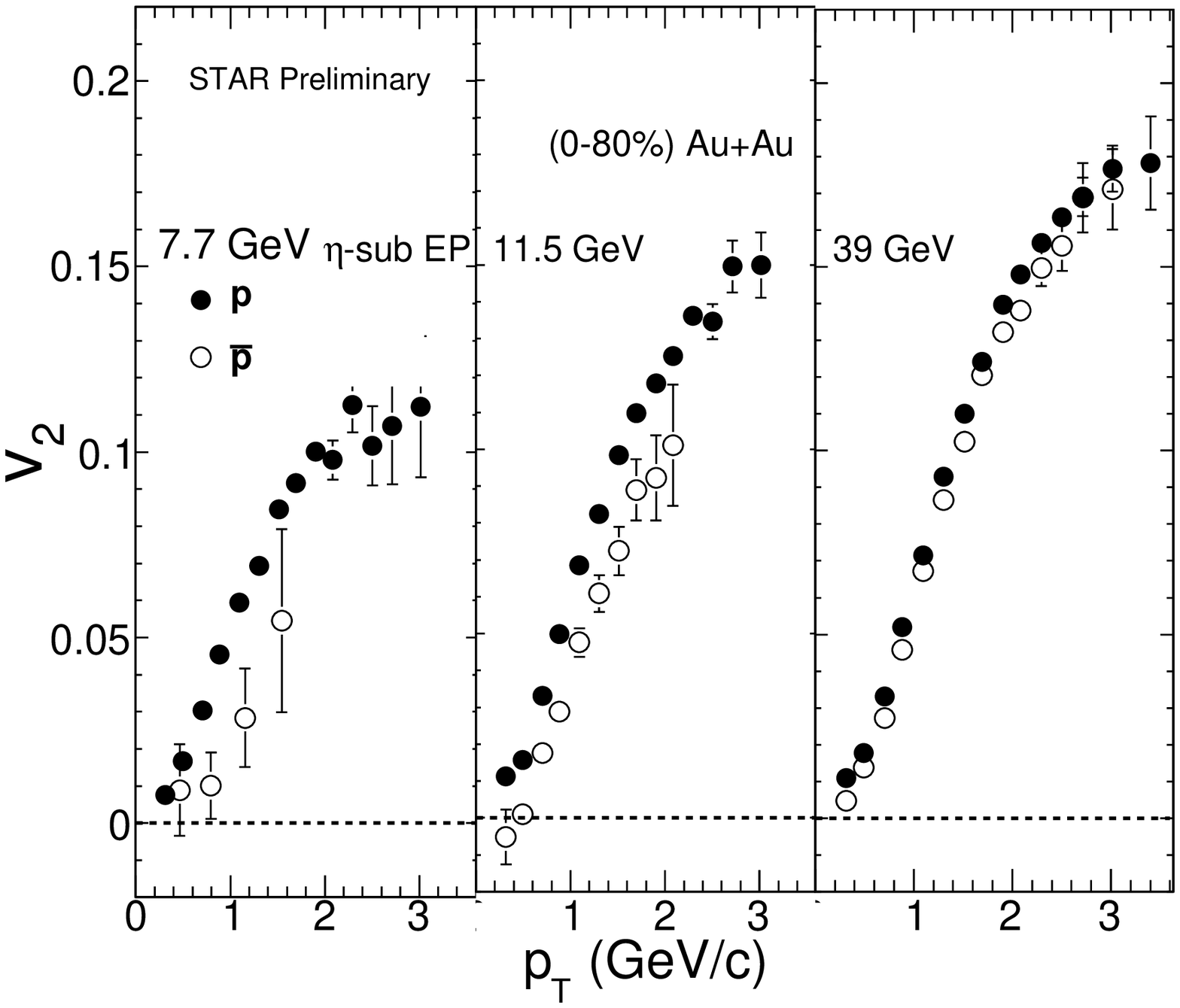}
\includegraphics[scale=0.3]{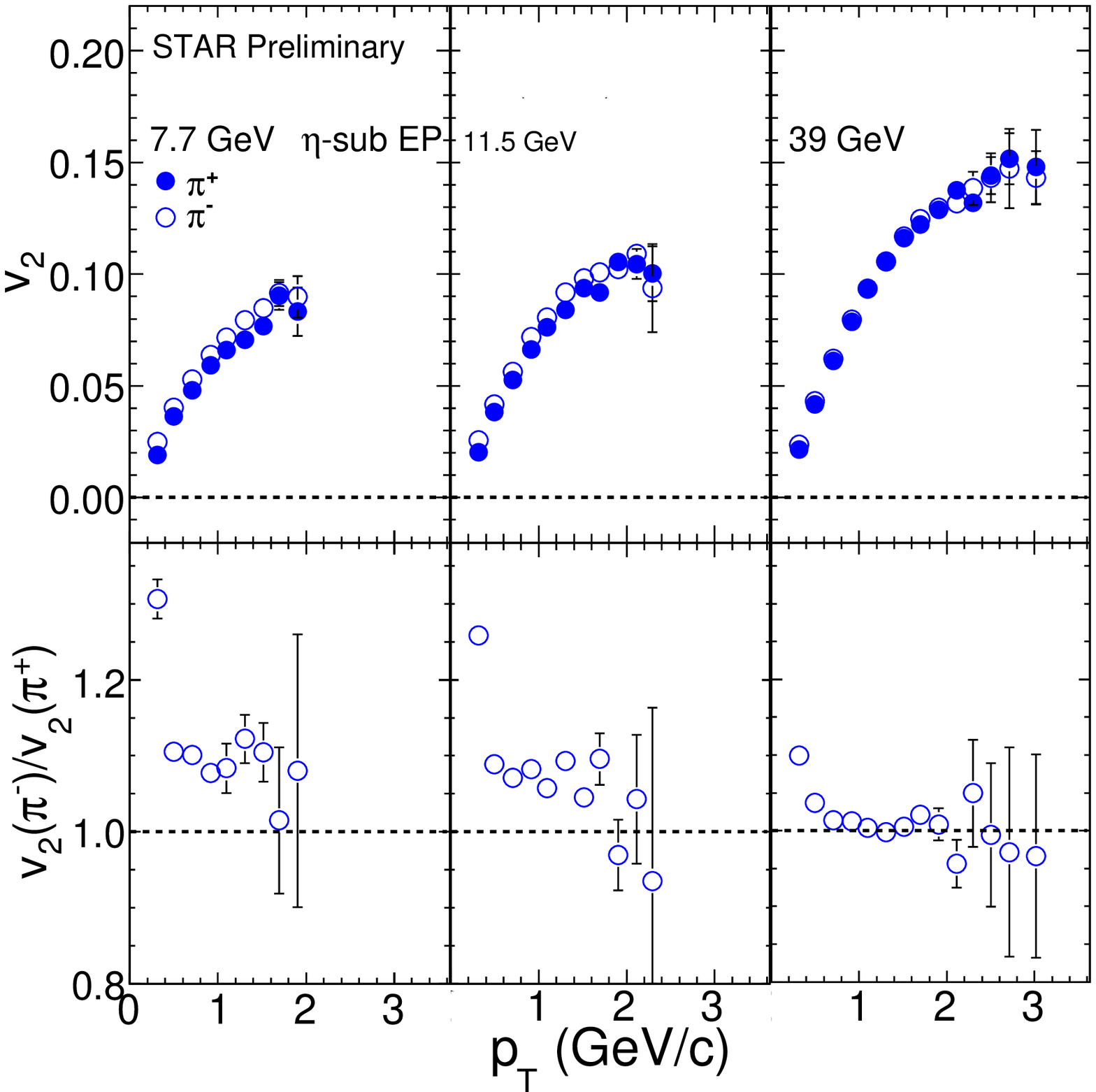}
\includegraphics[scale=0.3]{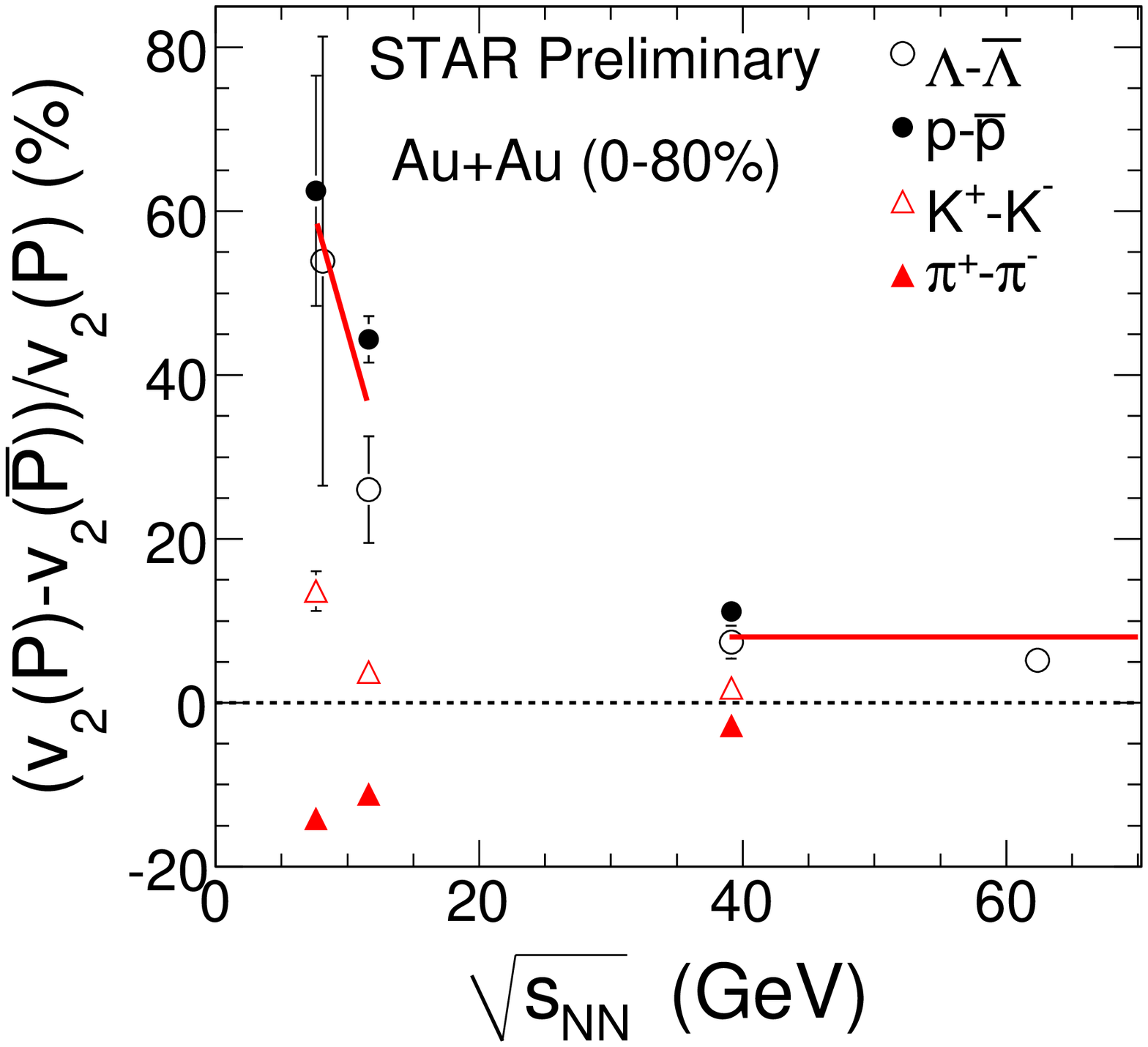}
\includegraphics[scale=0.3]{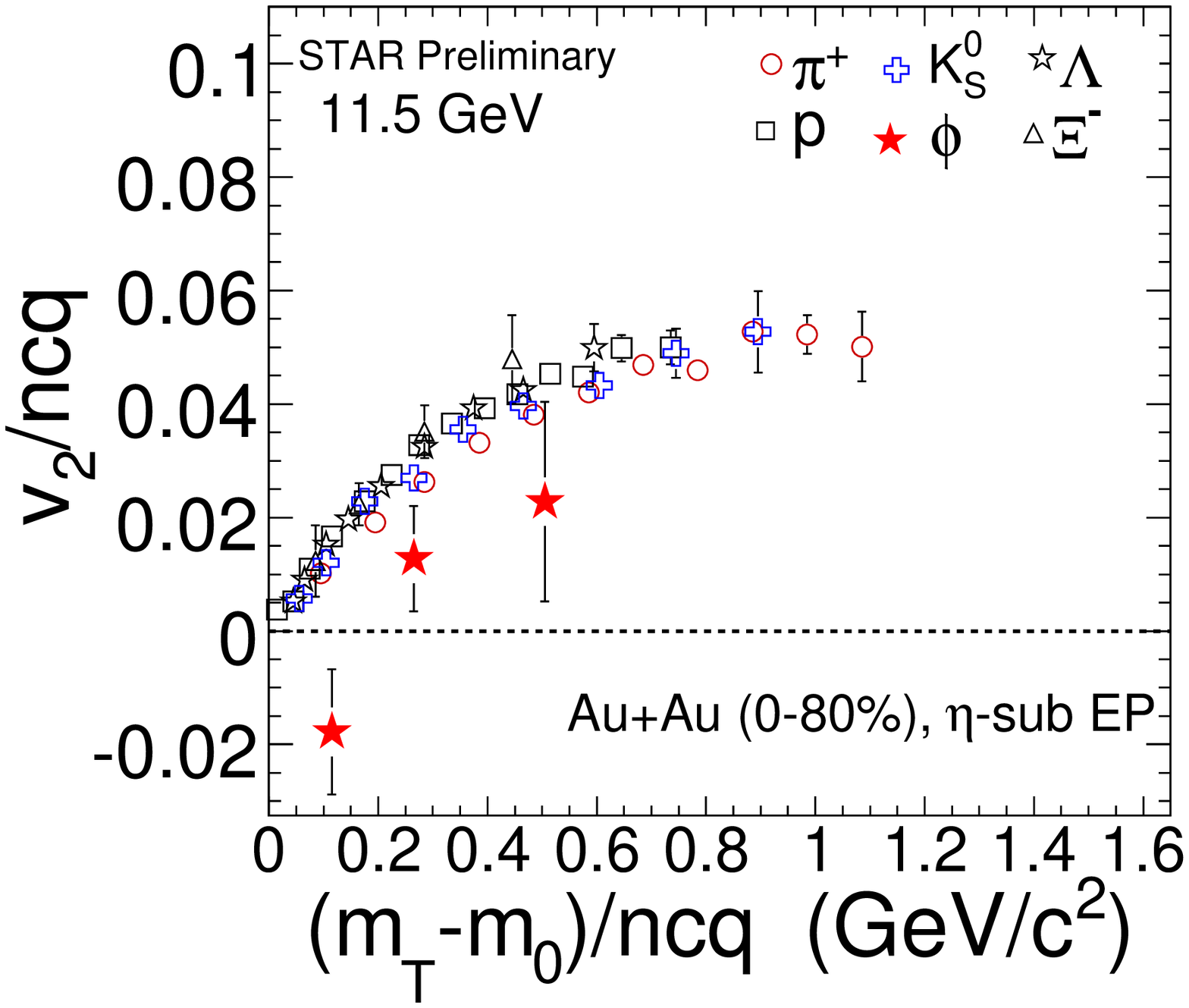}
\vspace{-0.5 cm}
\caption{(Color online) Top panel: $v_{2}$ of $p$, $\bar{p}$ and $\pi^{\pm}$
vs. $p_{T}$ in minimum bias Au+Au collisions at $\sqrt{s_{\mathrm {NN}}}$ = 7.7, 11.5, 39 GeV. 
Bottom left panel: Percentage difference between $v_{\rm 2}$ of particles and anti-particles. 
Bottom right panel: $v_{\rm 2}$/ncq vs. ($m_{\mathrm T} - m_{\rm 0}$)/ncq 
for 0-80\% Au+Au collisions at 11.5 GeV~\cite{alex}.}
\label{flow}
\end{center}
\end{figure}
The elliptic flow ($v_{2}$) is calculated as $\langle cos~2(\phi - \Psi_{\rm 2}) \rangle$, 
where $\phi$ denotes the azimuthal angle of the produced particles 
and $\Psi_{\rm 2}$ denotes the orientation of the second order event plane. 
Figure~\ref{flow} shows the $v_{2}$ of $p$, $\bar{p}$, $\pi^{+}$ and $\pi^{-}$
as a function of transverse momentum ($p_{\mathrm T}$) for minimum bias (0-80\%) 
Au+Au collisions at $\sqrt{s_{\mathrm {NN}}}$ = 7.7, 11.5 and 39 GeV~\cite{alex}. 
To avoid autocorrelations 
we calculated the event plane (EP) in two separated hemispheres 
with an $\eta$-gap of 0.05 on each side. Differences are observed in the $v_{\rm 2}$ of 
particles and anti-particles, which increases as the beam energy decreases. 
The percentage difference 
relative to $v_{\rm 2}$ 
of the particle is also shown in Fig.~\ref{flow} as a function of $\sqrt{s_{\mathrm {NN}}}$.
We find that $v_{\rm 2}$(anti-baryons) $<$ $v_{\rm 2}$(baryons), which could be due to
high net-baryon density matter at midrapidity. The $v_{\rm 2}(K^{-})$
$<$ $v_{\rm 2}(K^{+})$ for $\sqrt{s_{\mathrm {NN}}}$ = 7.7 and 11.5 GeV. This could be due to 
$K^{-}$ absorption in the medium, $K^{-}$--nucleon potential is attractive, or could
be related to dominance of associated production of kaons. The
$v_{\rm 2}(\pi^{-})$ $>$ $v_{\rm 2}(\pi^{+})$, this could be due the coulomb
repulsion of $\pi^{-}$ by the midrapidity protons or due to contributions from 
resonance decays or could be due to chiral magnetic effect~\cite{cme1}. 
The difference in particle and anti-particle $v_{\rm 2}$ also suggests that the 
number of constituent quark (ncq) scaling for all particle species 
(including nuclei) as observed at top RHIC energies~\cite{starflow} are no longer valid at 
these lower energies. Figure~\ref{flow} also shows the $v_{\rm 2}$
for identified particles scaled by ncq as a function of 
$(m_{\mathrm T} - m_{\rm 0})/ncq$.
Where $m_{\mathrm T}$ is the transverse mass of a hadron with mass $m_{\rm 0}$. We observe that $\phi$-meson
$v_{\rm 2}$ drops off the common trend followed by other particles. The interaction of
$\phi$-mesons with nucleons are expected to be smaller compared to other hadrons~\cite{bmnxu} 
and at top RHIC energy it was observed that $\phi$-mesons freeze-out close to the transition
temperature predicted by lattice QCD~\cite{starwp}.
All these make $\phi$-meson a carrier of early stage
information in heavy-ion collisions. $\phi$-meson $v_{\rm 2}$ was found to follow the ncq
scaling at top RHIC energy and this was used to conclude that a substantial amount of 
collectivity at 200 GeV has been developed at the partonic stage~\cite{starflow}. Small $\phi$-meson
$v_{\rm 2}$ at 11.5 GeV would then indicate collectivity contribution from partonic interactions 
to decrease with decrease in beam energy~\cite{bmnxu}.

\begin{figure}
\begin{center}
\includegraphics[scale=0.35]{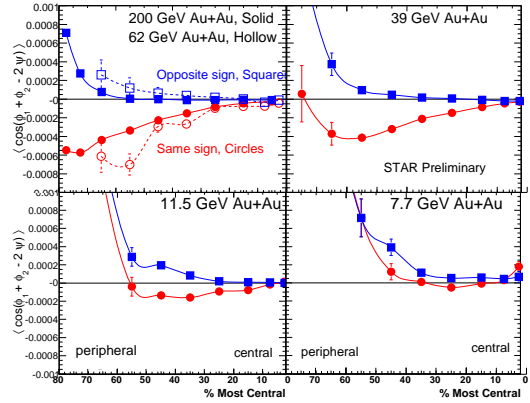}
\vspace{-0.5 cm}
\caption{(Color online) Charge hadron azimuthal correlations
with respect to reaction plane angle 
as a function of centrality for Au+Au collisions at midrapidity~\cite{dhevan}.} 
\label{lpv}
\end{center}
\end{figure}

\subsection{Dynamical charge correlations}
Figure~\ref{lpv} shows the results on charged hadron azimuthal correlations based on
3-particle correlation technique~\cite{starlpv}. The results are from Au+Au collisions 
at $\sqrt{s_{\mathrm {NN}}}$ = 7.7,11.5, 39, 64.2 and 200 GeV at midrapidity
between same charged and opposite charged hadrons 
with respect to reaction plane angle ($\Psi$)~\cite{dhevan}.
The observable, $\langle \cos(\phi_1 +\phi_2 -2\psi) \rangle$ 
represents the difference between azimuthal correlations projected
onto the direction of the angular momentum vector and correlations projected
onto the collision event plane. 
The difference between the same charge and opposite charge correlations 
at the higher energies seems to be consistent with the predictions for
existence of metastable domains in QCD vacuum leading to local {\it Parity} violation.
This phenomena needs deconfinement and chiral phase transitions which are expected to be
achieved in heavy-ion collisions (also referred as Chiral Magnetic Effect (CME))~\cite{fukushima}. 
We observe that the difference 
in correlations between same and opposite charges seems to decrease as beam energy
decreases and almost vanishes at 7.7 GeV. If the differences can be attributed
to QCD transitions, absence of it may indicate absence of such transitions
at the lower energies. The observable presented is parity-even, making it susceptible 
to physical processes not related to CME. In this respect STAR has carried 
out investigations with a new correlator which could be less sensitive to 
parity-even effects, the results of which suggest
that the measurements are consistent with CME~\cite{dhevan}. In addition, alternate 
underlying physics mechanisms are also investigated, which includes 
local charge conservations and flow~\cite{hui}. In a separate analysis using charge multiplicity asymmetry 
correlations, it was found that the charge separation is mainly present 
in the vicinity of the event-plane, and appears proportional to the 
particle distribution ellipticity~\cite{quan}.
Preliminary results indicate that the observed dynamical correlations could have
significant contributions from above sources as well.

\section{Search for signatures of softening of EOS}

\subsection{Freeze-out eccentricity}

\begin{figure}
\begin{center}
\includegraphics[scale=0.3]{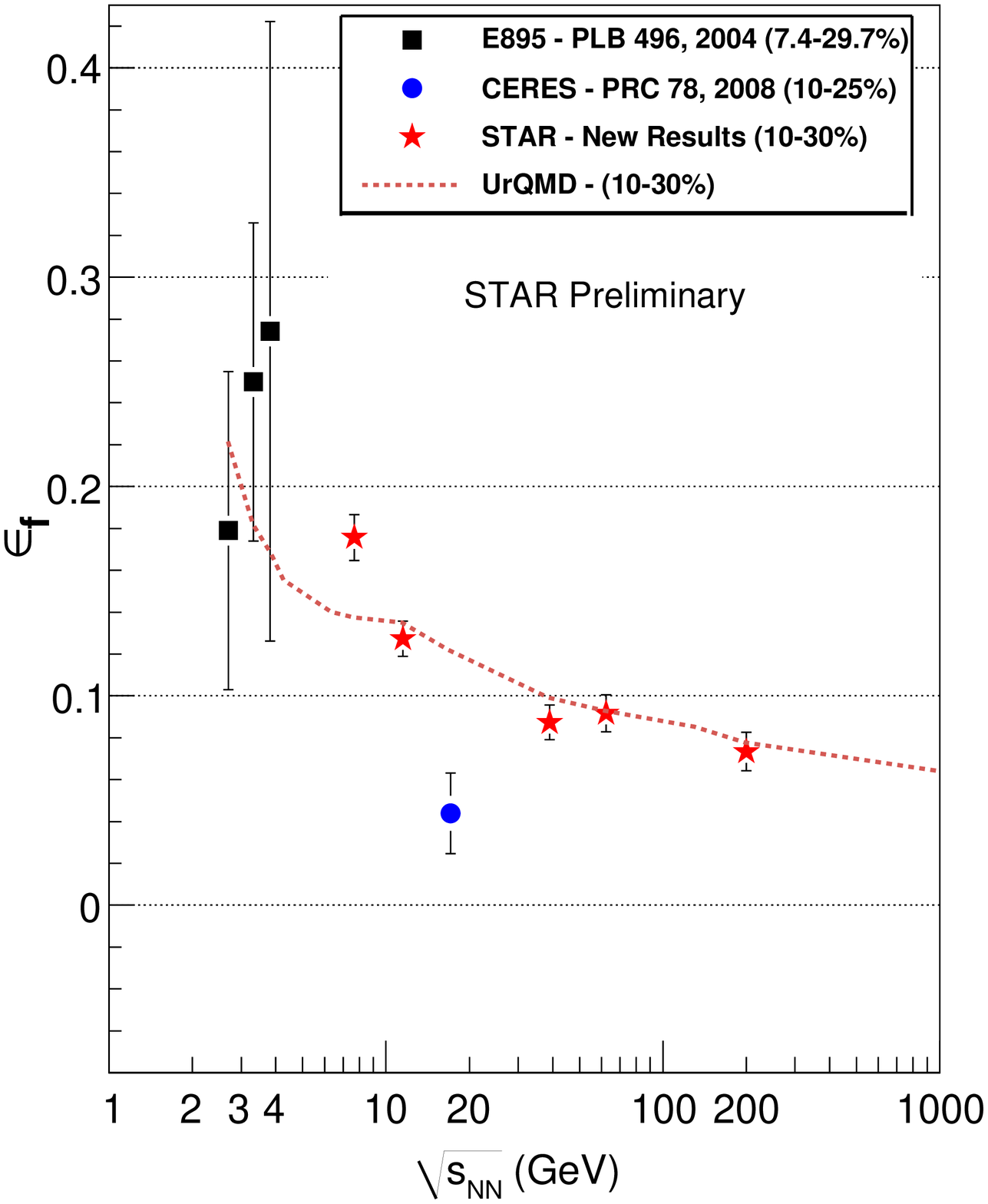}
\includegraphics[scale=0.3]{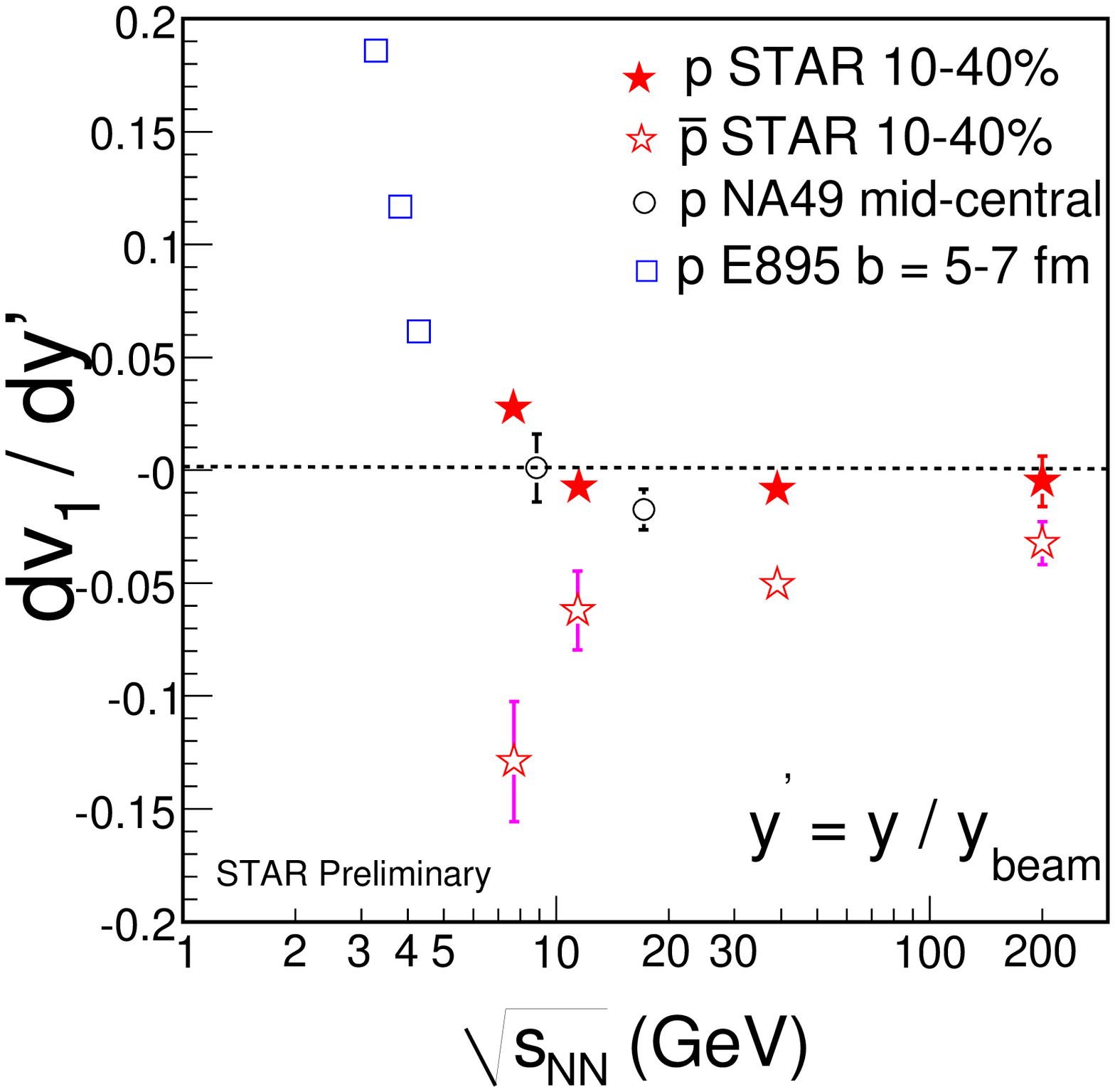}
\vspace{-0.5 cm}
\caption{(Color online) Left panel: Eccentricity vs.
$\sqrt{s_{\mathrm {NN}}}$ for mid-central heavy ion collisions
from AGS, SPS and RHIC are shown~\cite{chris}. The dashed lines are UrQMD model
calculations. Right panel: Slope of $p$ and $\bar{p}$ $v_{1}$ 
with respect to rapidity ($y$) normalized to beam rapidity vs. $\sqrt{s_{\mathrm {NN}}}$.}
\label{hbt}
\end{center}
\end{figure}

The eccentricity at the freeze-out ($\epsilon_{\mathrm f}$) is defined as the 
ratio $\frac{\sigma_{\mathrm y}^{2} - \sigma_{\mathrm x}^{2}}{\sigma_{\mathrm y}^{2} + \sigma_{\mathrm x}^{2}}$,
where $\sigma_{\mathrm x}$ and $\sigma_{\mathrm y}$ are the in- and out-of plane length
scales of the emitting source. These can be obtained by measuring the HBT
radii relative to the reaction plane in an event~\cite{chris, starhbt}. 
The $\epsilon_{\mathrm f}$ as a function of $\sqrt{s_{\mathrm {NN}}}$ is shown in Fig.~\ref{hbt} as
measured in AGS, SPS and RHIC experiments at midrapidity for mid-central 
collisions~\cite{chris}. Including the SPS result we observe a non-monotonic variation
of $\epsilon_{\mathrm f}$ with $\sqrt{s_{\mathrm {NN}}}$. RHIC data at 19.6
and 27 GeV this year may re-confirm this variation. Non-monotonic variations of freeze-out
volume (obtained using HBT observables) have been previously reported~\cite{cereshbt}
and interpreted as due to a constant mean free path at freeze-out 
and transition from a nucleon to a pion dominated freeze-out condition with
increase in $\sqrt{s_{\mathrm {NN}}}$. However $\epsilon_{\mathrm f}$ could also be sensitive
to the life time of the source and EOS for the system~\cite{hbttheory}.
In such a scenario, a non-monotonic variation of $\epsilon_{\mathrm f}$  vs. $\sqrt{s_{\mathrm {NN}}}$
could be considered as a signature of softening of the EOS~\cite{lisa}.

\subsection{Directed flow}
The directed flow ($v_{\rm 1}$) is calculated by computing the 
$\langle cos (\phi - \Psi_{\rm 1}) \rangle$, where $\Psi_{\rm 1}$ denotes the orientation 
of the first order reaction plane. The $v_{\rm 1}$($y$) reflects the 
collective side-ward motion of the particles in the final state. 
A negative slope of $v_{\rm 1}$($y$) for nucleons are expected to
be due to positive space-momentum correlations and baryon stopping. While specific
variations of $v_{\rm 1}$ with $y$ could also be sensitive to the softening of the
EOS~\cite{v1theory}. Figure~\ref{hbt} right panel shows the slope of $v_{\rm 1}$ for 
protons and anti-protons as a function of $\sqrt{s_{\mathrm {NN}}}$ for mid-central heavy-ion 
collisions~\cite{yadav}. 
The slope which describes the transverse side-ward motion of the particles 
relative to the beam direction is observed to change sign as the 
$\sqrt{s_{\mathrm {NN}}}$ decreases. The slope for $\bar{p}$ remains negative for $\sqrt{s_{\mathrm {NN}}}$ studied.
The difference in slope for $p$ and $\bar{p}$ are also observed to 
decrease with increase in $\sqrt{s_{\mathrm {NN}}}$. At a given BES energy,
the proton slope also changes sign as a function of collision centrality, 
being negative for peripheral collisions and positive for central collisions. 

\section{Search for signatures of critical point}

\begin{figure}
\begin{center}
\includegraphics[scale=0.3]{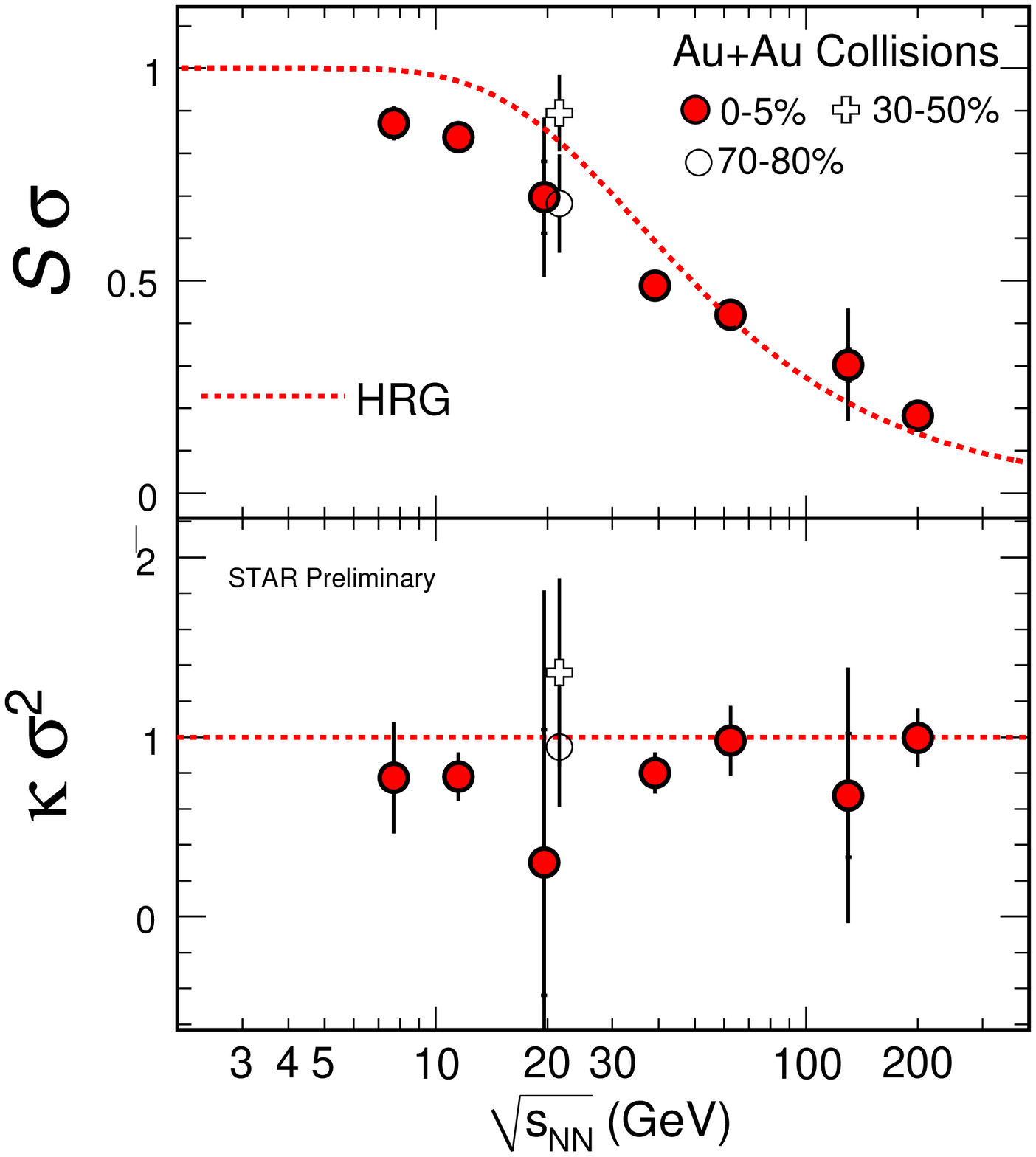}
\includegraphics[scale=0.3]{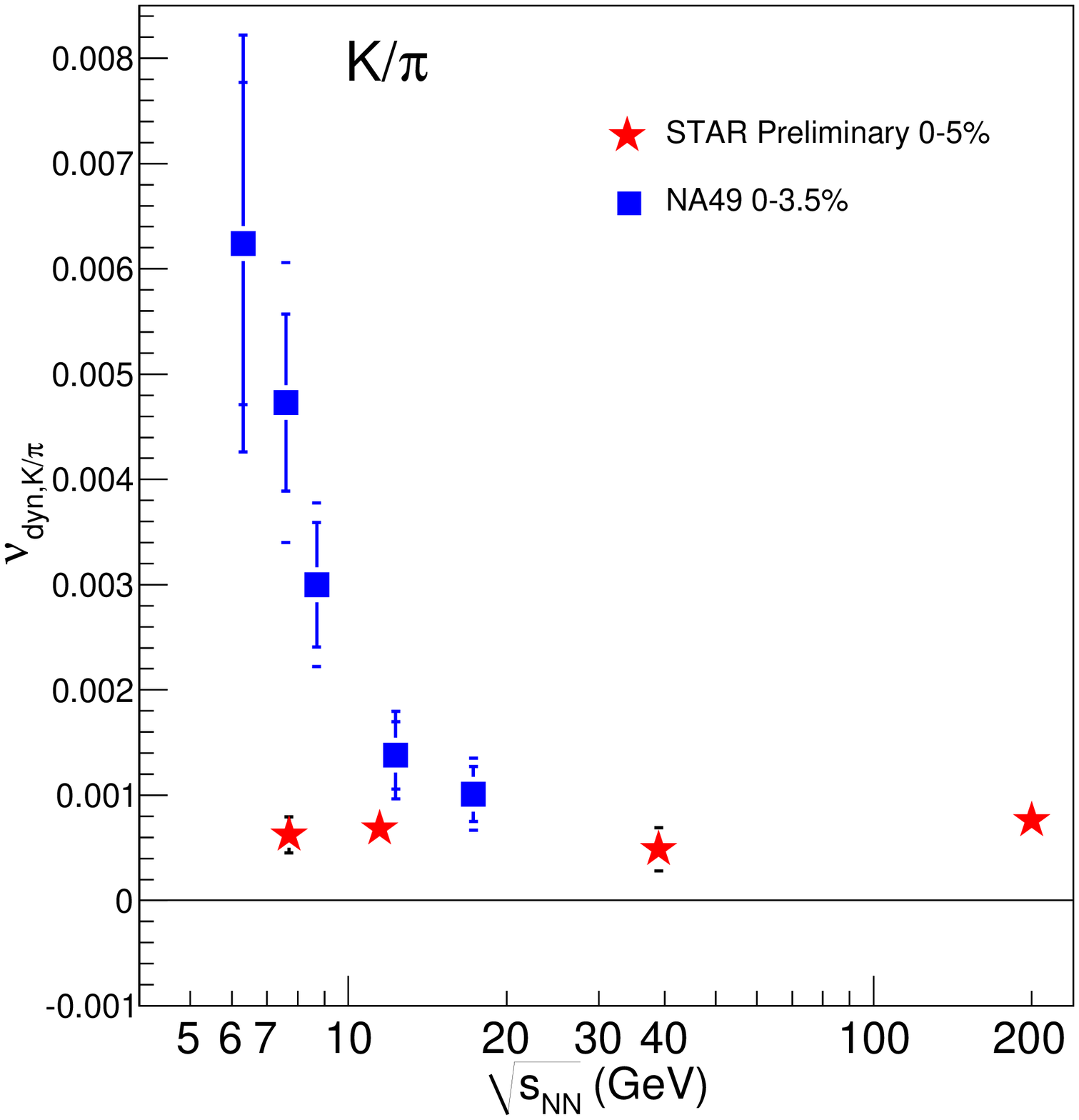}
\vspace{-0.5 cm}
\caption{(Color online) Left panel: $S\sigma$ and $\kappa\sigma^2$ 
vs. $\sqrt{s_{\mathrm {NN}}}$ for net-protons~\cite{luo}. The dashed line are
the Hadron Resonance Gas (HRG) model results~\cite{hrg}.
Right panel: $\nu_{\mathrm {dyn}}$($K/\pi$) vs. $\sqrt{s_{\mathrm {NN}}}$ for
Pb+Pb and Au+Au collisions at SPS and RHIC respectively.}
\label{fluc}
\end{center}
\end{figure}

\subsection{Higher moments of net-proton distributions}
In a static, infinite medium, the correlation length ($\xi$) diverges 
at the CP. 
Finite size and time effects in heavy-ion collisions put constraints on the 
values of $\xi$. The higher moments of distributions of conserved quantities
(net-baryons, net-charge, and net-strangeness), 
measuring deviations from a  Gaussian, have a sensitivity to CP fluctuations 
that is better than that of variance ($\sigma^2$), due to a stronger dependence 
on $\xi$. Recently it has also been shown that a crossing 
of the phase boundary can manifest itself by a change of sign of skewness ($\it{S}$) 
as a function of energy density and negative value of 
kurtosis ($\kappa$) could indicate existence of a CP in the vicinity~\cite{cp}. Further
higher moments are also proposed to be a sensitive probe for chiral phase transition
effects~\cite{hrg}. The products
$ S\sigma \sim \frac{\chi^{(3)}}{\chi^{(2)}}$ and 
$\kappa\sigma^2 \sim \frac{\chi^{(4)}}{\chi^{(2)}}$ are very useful
observables to measure as they cancel the volume effects and have
a direct connection to ratios of various order susceptibilities ($\chi$)
calculated using lattice QCD~\cite{science} and HRG model. 

Figure~\ref{fluc} shows $S\sigma$ and $\kappa\sigma^2$ versus $\sqrt{s_{\mathrm {NN}}}$ 
for net-protons (number of protons - number of anti-protons) distribution 
measured at midrapidity for the $0.4 < p_{\mathrm T} < 0.8$ GeV/$c$ in central Au+Au 
collisions. The results are compared to expectations from HRG model  
(similar values from Poissonian statistics). The results can test the predictions 
of non-perturbative QCD calculations at high temperature~\cite{science}.
The measurements show deviation from HRG for $\sqrt{s_{\mathrm {NN}}} < 39$ GeV. A non-monotonic variation
of $\kappa\sigma^2$ with $\sqrt{s_{\mathrm {NN}}}$ could indicate presence of CP.
The new data at $\sqrt{s_{\mathrm {NN}}}$ = 19.6 and 27 GeV will provide
a more complete picture on the presence of CP over a $\mu_{B}$ range
of 20 - 400 MeV. 

\subsection{Particle ratio fluctuations}

Particle ratio fluctuations are sensitive to particle numbers
at chemical freeze-out and may be insensitive to volume effects. A non-monotonic
variation of these fluctuations with $\sqrt{s_{\mathrm {NN}}}$ is expected
in presence of a CP. Figure~\ref{fluc}
right panel shows the dynamical fluctuations in $K/\pi$ ($\nu_{\mathrm {dyn}}$, which 
is zero for statistical fluctuations)~\cite{terry} measured by NA49 and 
STAR experiment as a function of $\sqrt{s_{\mathrm {NN}}}$.
The STAR results shows that the $\nu_{\mathrm {dyn}}$($K/\pi$) are constant
as a function of the $\sqrt{s_{\mathrm {NN}}}$ studied. Similar measurements have been carried 
out in STAR for fluctuations in $K/p$ and $p/\pi$ ratios. These exhibit a negative value of 
$\nu_{\mathrm {dyn}}$ indicating correlated emission of particles and have a monotonic 
variations with $\sqrt{s_{\mathrm {NN}}}$~\cite{terry}. Differences are observed 
between results from NA49 and STAR experiment in the common $\sqrt{s_{\mathrm {NN}}}$
region and for ratios involving kaons. This could be due to difference in
the experimental acceptances and the particle
identification techniques used. STAR  has a uniform $y$ vs. $p_{\mathrm T}$ acceptance 
for different particle species at midrapidity and uses TPC and TOF, 
to event-by-event identify the charged hadrons.  

\section{Summary}
The BES program has extended the $\mu_{\mathrm B}$ range at RHIC from 20 MeV to
about 400 MeV, thereby covering a large part of the conjectured QCD phase diagram. 
Several new observations are reported. The identified hadron production and 
freeze-out parameters revealed a high net-baryon density at midrapidity is achieved.
First observations of significant differences in particle and 
anti-particle $v_{\rm 2}$ could be attributed to such an effect. 
Hadronic interactions start to dominate over partonic interactions around 11.5 GeV. 
This is indicated by the measurement of small $v_{\rm 2}$ for the $\phi$-mesons
relative to other hadrons at 11.5 GeV and a reduction in the difference
in dynamical charge correlations between same and opposite charges with
respect to reaction plane. Interesting $\sqrt{s_{\mathrm {NN}}}$ dependence 
trends are observed in observables
which could be related to search for softening of the EOS. The freeze-out
eccentricity shows a possible change in slope around the BES energies. In addition 
the slope of proton $v_{\rm 1}$($y$) changes sign at these energies. 
STAR detector provides a large and uniform acceptance around midrapidity 
with excellent event-by-event particle identification capabilities to 
measure multiplicity fluctuations for CP search. Deviations of higher order fluctuations 
for the net-proton distributions from HRG expectations 
are observed at 7.7, 11.5 and 39 GeV. These measurements are being used
to study the structure of the QCD phase diagram. We expect to complete the first
phase of BES program before 2012. 

{\it Acknowledgments:}
This work is supported by DAE-BRNS project sanction No. 2010/21/15-BRNS/2026.

\end{document}